Check for updates

Cite this: DOI: 10.1039/c8sc01162a

# Shaping excitons in light-harvesting proteins through nanoplasmonics†

Stefano Caprasecca, [ID] *[a] Stefano Corni [ID] ‡*[b] and Benedetta Mennucci [ID] *[a]

Nanoplasmonics has been used to enhance molecular spectroscopic signals, with exquisite spatial resolution down to the sub-molecular scale. By means of a rigorous, state-of-the-art multiscale model based on a quantum chemical description, here we show that optimally tuned tip-shaped metal nanoparticles can selectively excite localized regions of typically coherent systems, eventually narrowing down to probing one single pigment. The well-known major light-harvesting complex LH2 of purple bacteria has been investigated because of its unique properties, as it presents both high and weak delocalization among subclusters of pigments. This finding opens the way to the direct spectroscopic investigation of quantum-based processes, such as the quantum diffusion of the excitation among the chromophores, and their external manipulation.

## Introduction

Nanoplasmonics is the primary toolbox to manipulate light at the nanoscale, i.e., in the realm of molecules, proving to be an extremely flexible and powerful tool both to control photochemical processes and to enhance optical responses of (supra) molecular systems.[1–7] One of its earlier successes was to scale the sensitivity of spectroscopic techniques down to the single molecule level, in Raman scattering[8,9] and also in fluorescence.[10,11] More recently, the enhanced molecular spectroscopic signals allowed to probe even the sub-molecular scale.[12] The possibility of reaching the strong-coupling regime[13–16] has been also demonstrated for a single molecule.[17] Such achievements, in particular, were made possible by the spatially inhomogeneous nature of the plasmon-enhanced electromagnetic field,[18,19] that can even be stretched down to the atomic scale.[20] In parallel, it has been shown that nanoplasmonics can be joined with ultrafast spectroscopy, to endow nanoplasmonics with time resolution.[21–24] These works taken together demonstrate the potential of nanoplasmonics in investigating the intimate nature of molecular quantum phenomena, far beyond what can be achieved by far field spectroscopy.

Among the many phenomena that can be studied, one of the most debated topics in the last ten years has been the role of quantum coherence effects in the light-harvesting function of photosynthetic organisms.[26–29] Typical pigment–protein complexes devoted to this task contain tens of chromophores that sustain several delocalized excitations (excitons). In this context, nanoplasmonics has been used in combination with advanced optical spectroscopy to investigate the possibility of controlling the exciton properties of the complexes,[30–32] in some cases achieving a single molecule sensitivity.[33–35] Although ultrafast spectroscopy is the best tool available for the time being, it is also rather limited in terms of the exciton states that can be created and manipulated, due to the well-known dipolar light–matter interaction selection rules. It is very challenging, for example, to directly test to which extent a quantum walk model for the exciton diffusion within the chromophoric system is appropriate, i.e., how important quantum coherences really are.[28] In fact, the ideal experiment would be to excite a single (or a few) chromophore(s) inside the protein and then follow the resulting evolution of the quantum state of the system. But the excitation of a single chromophore, whose states are hybridized with those of all the others, cannot be achieved with a far field excitation only.

By means of a rigorous, state-of-the-art multiscale model[36,37] which combines time-dependent density functional theory (TDDFT) and two polarizable classical models for the protein and the metal nanoparticle (MNP), we show that the spatially inhomogeneous electromagnetic field provided by sharp metal tips can be used to create localized excitations in an otherwise delocalized multichromophoric system. The system under study, LH2, is a light-harvesting pigment–protein complex present in purple bacteria, comprising 27 bacteriochlorophyll-a units (BChl) arranged in two rings, labeled B800 and B850, as shown in Fig. 1A. The light absorption of the complex in the 750–900 nm range is determined by the excitonic interactions of the $Q_y$ transition in each BChl: the resulting spectrum is in fact characterized by clear excitonic features, with two distinctive bands at ~800 and ~850 nm, that are spatially delocalized over the B800 and B850 rings, respectively (see Fig. 1B). Two tip-shaped gold nanoparticles have been designed to display plasmon peaks in resonance with the two absorption bands of LH2, at 800 and 850 nm, as in Fig. 1E, and in order to enhance their effect on the BChl units, the former was placed in correspondence to the B800 ring, while the latter in correspondence to the B850 ring.

[a]Dipartimento di Chimica e Chimica Industriale, Università di Pisa, I-56124 Pisa, Italy. E-mail: stefano.caprasecca@for.unipi.it; benedetta.mennucci@unipi.it

[b]Dipartimento di Scienze Chimiche, Università di Padova, I-35131 Padova, Italy. E-mail: stefano.corni@unipd.it

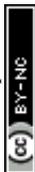

While the effects of metal proximity minimally perturbs the inter-chromophoric energy transfer rates, the tip-enhanced field is inhomogeneous enough to localize a pulse excitation on a few chromophores. Such localized excitation remains possible also when thermal fluctuations are accounted for. These findings open the way to the direct investigation of quantum-based processes such as the quantum diffusion of the excitation among the chromophores, rather than inferring them from the far-field behavior of the ultrafast response of the system. Also, they show that exciton states can be manipulated and shaped from outside, without the need of mutating the supramolecular system.

## Methods

The structure of LH2 is taken from *Rhodopseudomonas acidophila* (PDB: 1NKZ, resolution 2 Å).[38] The LH2 system is oriented with the $C_9$ axis along $z$. There are 18 BChls in the B850 ring, labeled $\alpha$ and $\beta$ alternately, and 9 BChls in the B800 ring, labeled $\gamma$ (see Fig. 1C). The distinction between $\alpha$ and $\beta$ BChls is due to the different apoprotein chains they are noncovalently bound to. The crystal structure is refined by adding hydrogens, and considering all titrable residues in their standard protonation state. Histidine residues coordinated to bacteriochlorophylls are protonated at the $\delta$ position to allow coordination with the Mg atoms. The positions of the hydrogen atoms are optimized at the Molecular Mechanics (MM) level using Amber14.[39] The geometries of the three non-equivalent bacteriochlorophylls are optimized at the B3LYP/6-31G(d) level within a fixed protein environment, using the ONIOM (our own n-layered integrated molecular orbital and molecular mechanics) scheme.[40] The geometries thus optimized are replicated exploiting the $C_9$ symmetry of LH2, in order to obtain a completely symmetrical complex. The geometry of BChl is also optimized in acetone solution with PCM.

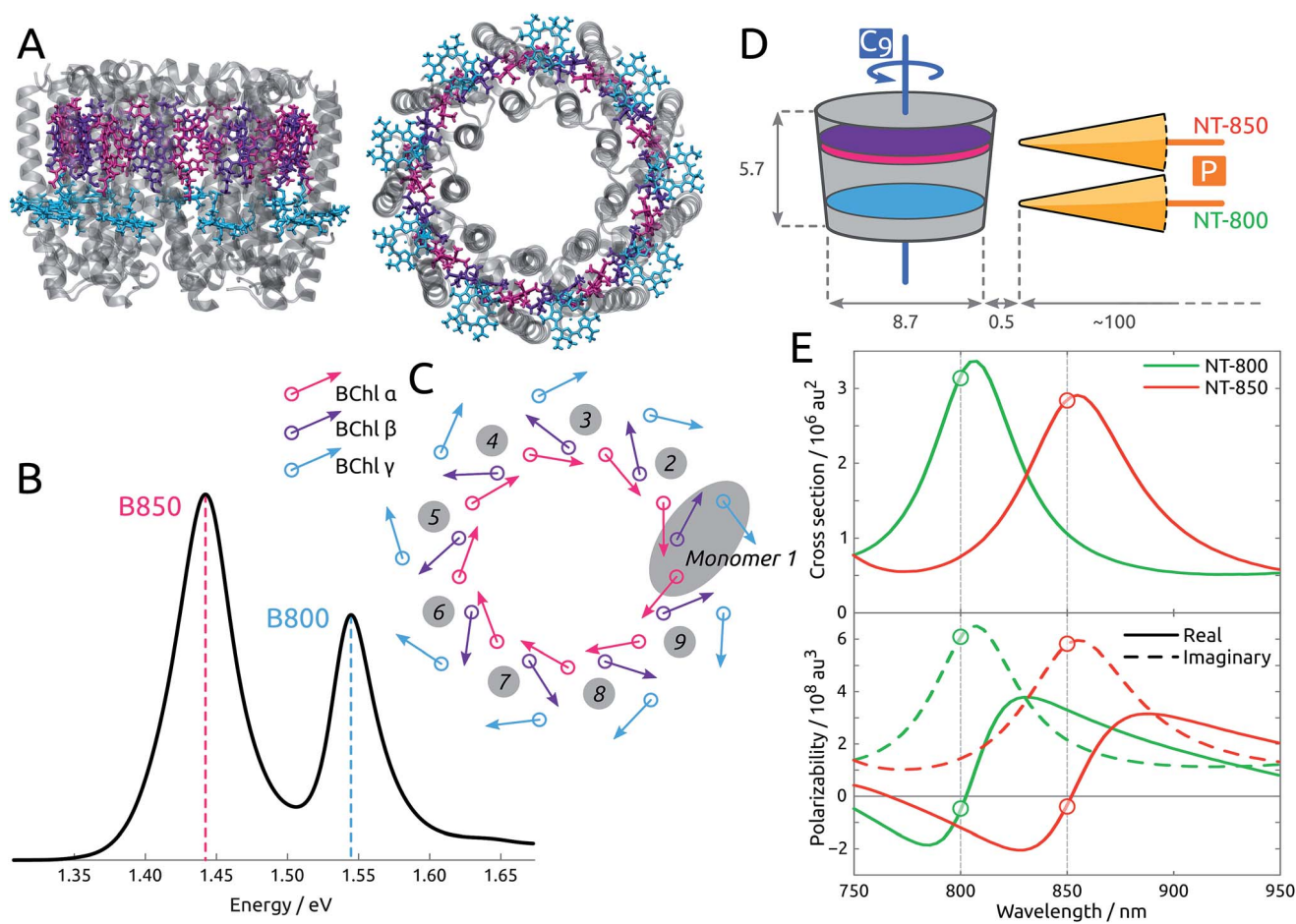

Fig. 1 (A) Side and top view of the LH2 complex. The B850 BChls are divided into $\alpha$ (pink) and $\beta$ (purple) due to the different local environment. The B800 BChls, also labeled $\gamma$, are colored cyan. (B) Experimental absorption spectrum,[25] measured at 300 K. (C) Schematic representation of the arrangement of the transition dipole moments corresponding to the first bright excited state ($Q_y$ state) of each BChl in the complex. (D) Tips NT-800 and NT-850 are placed with the polarization axis (labeled P) perpendicular to the LH2 $C_9$ axis. Dimensions and distances are reported in nm. (E) (Top) Simulated extinction cross sections for gold NT-800 and NT-850. (Bottom) Simulated metal polarizability along the main axis for the same tips, showing real and imaginary components separately.

Each of the 27 BChl units is treated as a two-level system, as only the bright $Q_y$ transition is considered. Calculations of BChl transition energies and properties are carried out at TD-DFT level with the CAM-B3LYP functional and the 6-31G(d) basis set, as in previous works of some of us on LH2.[41]

The protein environment is included using a classical atomistic representation where the atoms are described using fixed point charges and atomic polarizabilities (MMPol).[42,43] Within this approach, the environment is able to polarize as a response to the quantum-mechanical electron density. This is of fundamental importance, especially when electronic excitations are studied. The parameters of the classical atoms are as described elsewhere in previous works of ours.[37,41]

The model used to account for the plasmonic effects is an extension of the Polarizable Continuum Model (PCM) originally developed to describe solvents.[44] The metal nanoparticle (MNP) is represented as a continuous body characterized by its response properties to electric fields (both the external ones and those generated by the LH2 charge distribution): a conductor for static perturbations and a dielectric characterized by a frequency-dependent complex permittivity for dynamic perturbations. The response of the MNP is described in terms of induced charges spread on its surface, which is built using interlocked spheres of different radii and discretized by using a triangular mesh. For more details on the theory and implementation we refer to the literature.[36,45–47]

Two gold nanotips (NTs) are employed, whose shape and dimension are shown in Fig. S3 of the ESI.† The shape and size of the nanotips are compatible with those of experimentally synthesized nanoparticles,[48] and have pointy ends that correspond to those inducing the experimental results observed in ref. 20. Some rod-shaped nanoparticles, lacking pointy ends, have also been tested, but yielded less relevant results, as they affected the exciton properties of the pigment–protein complex less strongly, as already reported,[37] and will not be further analyzed here. In all cases, the experimental frequency-dependent complex permittivity of gold is used.[49] The tips are designed to display plasmon peaks in resonance with the two absorption bands of LH2, at 800 and 850 nm, as in Fig. 1E, and are labeled NT-800 and NT-850 accordingly. They have similar shapes, characterized by a 1 nm-large tip and ∼100 nm length, and are placed with their polarization axis (P) perpendicular to the LH2 $C_9$ axis, at a very short distance from the protein, in correspondence to the BChl rings, as in Fig. 1D. To enhance their effect on the BChl units, NT-800 is placed in correspondence to the B800 ring, while NT-850 is placed in correspondence to the B850 ring. The closest BChls are $\gamma_1$ and $\gamma_9$ (NT-800) and $\alpha_1$ and $\beta_1$ (NT-850), at a distance of ∼0.5 nm.

The exciton Hamiltonian reads

$$H = \sum_n \varepsilon_n |n\rangle \langle n| + \sum_{m \neq n} V_{m,n} |m\rangle \langle n| \qquad (1)$$

where $\varepsilon_n$ is the $Q_y$ excitation energy (site energy) localized on the $n$-th BChl and $V_{m,n}$ is its electronic coupling with the corresponding excitation on the $m$-th BChl. Here the lower-case indices are used to label localized states. The electronic couplings are computed from the Coulomb interaction between transition densities corresponding to the $Q_y$ excitations on different BChls. The explicit effects of the plasmonic and the protein environment are also included as two additional terms which depend on the dipole moments induced by the $Q_y$ transitions on the protein atoms[42] and on the charges induced by the same transition on the MNP surface.[50] Due to the weak coupling between the two rings, the B800 and B850 blocks of the Hamiltonian are separately diagonalized, resulting in exciton states fully placed over either of the two rings.[37] The eigenvalues and eigenvectors obtained from the diagonalization correspond to the exciton energies $\{E_M\}$ and exciton states $\{|M\rangle\}$ (uppercase indices label exciton states). The eigenvectors form the coefficient matrix **C**, whose elements $C_{Mm}$ identify the contribution of the $m$-th BChl to the $M$-th exciton state. The extent of delocalization of the exciton states is quantified in terms of the participation ratio which is defined as[51]

$$\mathrm{PR}_M = \frac{1}{\sum_m C_{Mm}^4} \qquad (2)$$

This definition ensures that $PR \in [1, N_M]$, where $N_M$ is the number of localized states $\{|m\rangle\}$ of which the exciton states are a combination. In the present work, PR values are between 1 and 9 (B800 exciton states) and between 1 and 18 (B850 states).

The state excited after irradiation with light in a certain frequency range $\Omega$ is a superposition of exciton states whose energy lies in that range, weighted by the component of their transition dipole moment along the direction of the light polarization $\hat{\varepsilon}$:

$$|\Psi\rangle = \frac{1}{\mathscr{N}} \sum_M z_M |M\rangle \qquad (3)$$

where $z_M = \hat{\varepsilon} \cdot \mu_M^{\mathrm{tr}}$. Here $\mu_M^{\mathrm{tr}}$ is the exciton transition dipole moment, defined in terms of the localized ones, $\mu_m^{\mathrm{tr}}$, where the transition considered is that between the ground and first excited state of BChl: $\mu_M^{\mathrm{tr}} = \sum_m C_{Mm} \mu_m^{\mathrm{tr}}$. Recall that $z_M$ in eqn (3) are complex since the effective dipole moment is a complex vector. The normalization constant is $\mathscr{N} = \sqrt{\sum_M |z_M|^2}$. The exciton states can be further expanded in the basis of localized transitions $\{|m\rangle\}$ through the coefficient matrix **C**:

$$|\Psi\rangle = \frac{1}{\mathscr{N}} \sum_M \sum_m z_M C_{Mm} |m\rangle \qquad (4)$$

The projection of $|\Psi\rangle$ along one localized state $|q\rangle$ therefore is:

$$\langle q|\Psi\rangle = \frac{1}{\mathcal{N}}\sum_M z_M C_{Mq} \qquad (5)$$

since the localized states are assumed orthonormal. The population of state $|q\rangle$ is the square modulus of the projection:

$$P_q = |\langle q|\Psi\rangle|^2 = \frac{\left|\sum_M z_M C_{Mq}\right|^2}{\sum_M |z_M|^2} \qquad (6)$$

The spectra in the absence of any metal nanoparticle are computed using the unperturbed coefficient matrix $\mathbf{C}_o$ and unperturbed transition dipole moments $\mu_o^{tr}$ (comprising both the BChl transition dipoles, $\mu_{sys}^{tr}$, and those induced on the polarizable protein by the former ones, $\mu_{prot}^{tr}$). To account for the static disorder in the spectra, a Gaussian distribution of site energies with standard deviation 40 and 250 cm$^{-1}$ is used for B800 and B850 BChls, respectively (data from ref. 52). For each realization of the disorder, a spectrum is simulated using Lorentzian lineshapes at the transition energies with 5 cm$^{-1}$ width. The 100 000 realizations are then averaged. All the other spectra presented are broadened artificially using Lorentzian lineshapes of HWHM 200 and 160 cm$^{-1}$ for exciton states on the B850 and B800 rings, respectively.

The close presence of a MNP has two concurring effects on the optical properties of the aggregate, which will be referred to as intrinsic and extrinsic.

The intrinsic effect consists of the following: the metal perturbs the exciton states of the LH2 complex by acting on the elements that make up the exciton Hamiltonian of eqn (1). Both site energies and couplings are computed in the presence of the composite and polarizable environment. In particular, the presence of the MNP not only perturbs the Coulomb-dominated coupling between BChls, but also introduces a second-order, polarization-mediated complex interaction term.[50,53] The combined effect on $\varepsilon$ and $V$ alters the exciton properties of the complex, and is due both to the protein environment and the MNP. However, while the protein does not perturb significantly the symmetry of the system, as its effect is similar for analogous BChls, the nanoparticle, positioned asymmetrically with respect to the $C_9$ axis, can generally have a much stronger effect. The spectra including the "intrinsic" effect of the MNP are computed using the perturbed coefficient matrix $\mathbf{C}_{MNP}$ and the unperturbed transition dipole moments $\mu_o^{tr}$.

The presence of the metal also implies the onset of a larger effect, here referred to as extrinsic effect, which is related to the ability of the nanoparticle to affect the interaction between the molecular system and the light. When photon absorption is concerned, the absorption is proportional to the electric field intensity acting locally on the molecule. Due to the plasmonic resonance of the MNP, such field will be enhanced but also its spatial distribution will be modified becoming inhomogeneous. The spectra including the "extrinsic" effect of the MNP are computed using the perturbed coefficient matrix $\mathbf{C}_{MNP}$ and the perturbed transition dipole moments $\mu_o^{tr} + \mu_{met}^{tr}$.

We consider unpolarized light for all the main text results (some with polarized light are in the ESI†). We stress that there is a practical difference in the meaning of unpolarized light when we investigate the intrinsic effect only and when we investigate instead intrinsic + extrinsic effects. In the latter case, which is the realistic one, we refer to using unpolarized light to excite the tip. Due to the tip plasmonic response, the resulting field at the protein position is actually strongly polarized along the nanotip axis direction even if unpolarized light was used. When we focus instead on intrinsic effects only, which is a purely modelistic situation, the effects of the tip on light are purposely disregarded, and the unpolarized light is thus that directly acting on the protein.

A locally modified version of the Gaussian09 code[54] has been used in all calculations.

## Results and discussion

As detailed in Methods section, the presence of a MNP has two concurring effects on the optical properties of the aggregate, which will be referred to as intrinsic and extrinsic and will be analyzed separately.

### Intrinsic effect: modification of the exciton states

To investigate the intrinsic effect, we analyze the local modifications induced by the tip on the individual BChls, and the ensuing perturbation on the expansion coefficients from the BChl-localized basis to the exciton one, which are obtained by diagonalizing the exciton Hamiltonian. A perturbation of such coefficients alters all the exciton properties.

The tip NT-800 is placed with its polarization axis perpendicular to the LH2 $C_9$ axis and pointing to the B800 ring, as indicated schematically in Fig. 1D. As expected, a rather strong effect is observed on the excitation energy of the closest BChl unit, $\gamma_1$, which is blue-shifted by nearly 100 cm$^{-1}$ (see ESI Fig. S5†). As the other BChls are only minimally affected, the interaction is markedly local and potentially leading to an intense modification of the exciton states. Concurrently, the electronic couplings are also affected, although by a lesser extent. Notably, the couplings $V(\gamma_1, \gamma_2)$ and $V(\gamma_1, \gamma_9)$ are those of most strong concern, with their real components respectively decreasing and increasing by ~60% and ~40%. In particular, the predominant effect of the tip is through the complex polarization-mediated interaction term. However, the shift induced on the couplings is at most 25% of that on the site energies, and the net effect is therefore mainly governed by the latter.

Moving to the delocalized exciton basis, we observe that most exciton energies are only minimally affected, while significant blue shifts are induced on a few states (see ESI Table S2†). When inspecting the spatial properties of the exciton states, significant modifications appear in the presence of NT-800, particularly for states S10 and S22. The spatial distributions of these states are analyzed in Fig. 2, where their

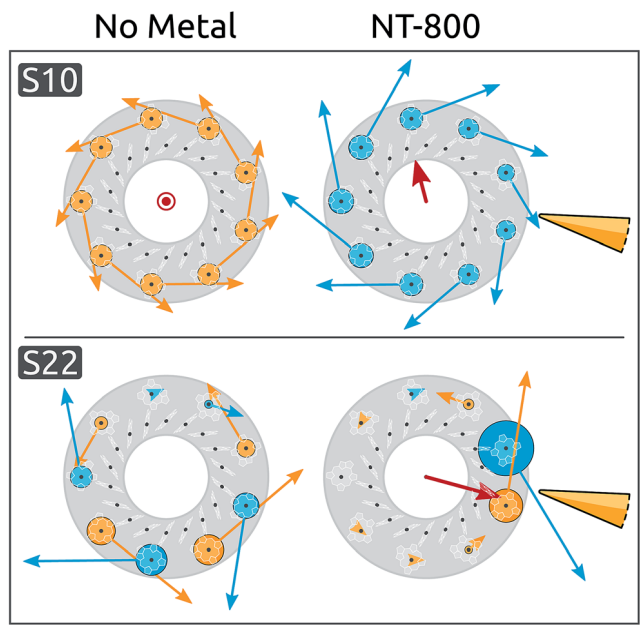

**No Metal**    **NT-800**

Fig. 2 Exciton coefficients and transition dipole moments of exciton states S10 (top) and S22 (bottom), without and with nanoparticle NT-800. The circles placed on the BChl units have radii proportional to the corresponding coefficients, and colored to indicate the sign. The arrows represent the localized transition dipole moments, weighted by their relative coefficients. The red arrows represent the resulting exciton transition dipole moments (real component only). In the NT-800 arrangement, the tip axis is lying on the BChl ring plane, perpendicular to the $C_9$ axis, and closest to the BChls $\gamma_1$ and $\gamma_9$, as indicated by the yellow tip. Drawing out of scale.

expansion coefficients are represented in the two cases without metal (left), and with NT-800 (right).

Initially, S10 is completely delocalized over the B800 ring and characterized by a very small dipole moment in the $C_9$ direction. The presence of the tip introduces a perturbation to the coefficient matrix which breaks the symmetry, decreasing the contribution from the closest BChls. As a consequence, the resulting exciton transition dipole moment, which is the weighted average of the localized dipoles, increases by ∼20% and gains a component perpendicular to the $C_9$ axis. This effect is even more evident for the S22 state, which is initially dark and rather delocalized on the B800 ring, and is subject to a strong localization over the $\gamma_1$ and $\gamma_9$ BChls, with a consequent increase of the dipole moment (from 0 to 1.9 a.u.). The level of delocalization can be quantified through the participation ratio (eqn (2)). By definition, the PR ranges from 1 (for a state fully localized on one chromophore) to $N$, the number of chromophores (for perfectly delocalized states). The PR reflects the metal-induced localization of S10 and S22: the former slightly reduces from the initial value of 9.0 (i.e., complete delocalization over the B800 ring) to 8.6, while the latter PR goes from 6.0 to 1.8 (see ESI Table S4†).

Such asymmetric redistribution of the states also reflects on the spectrum: in the absence of MNPs, the absorption spectrum is dominated by two pairs of very bright "fingerprint" exciton states, S2 and S3, at lower energy, and S11 and S12, at higher energy, that contain nearly all the oscillator strength of the B850 and B800 BChls, respectively. All the other exciton states are either dark or nearly dark by symmetry. As detailed above, the intrinsic effect of the tip is visible on both bright and dark states: on the one hand, it is able to remove the degeneracy in the dipole moments of the high-energy pair of bright states (see Fig. 3B and ESI Table S1†). At the same time, it causes a redistribution of the dipole moment between states in the same energy range, whence states initially dark gain strength.

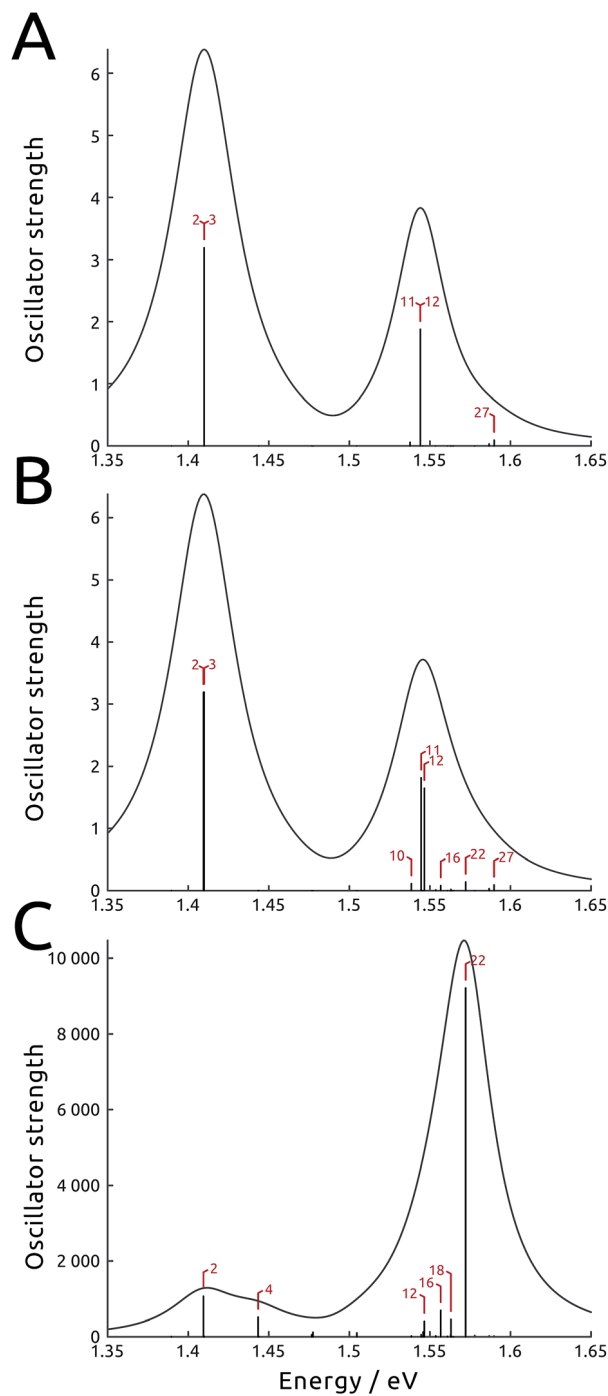

Fig. 3 Simulated absorption spectra of LH2 with and without NT-800. The exciton states responsible for the peaks are indicated. Broadening is simulated using Lorentzian lineshapes. (A) No metal is present. (B) The intrinsic effect of the metal on the exciton states is considered, but without including the explicit contribution of the NT to the transition dipole. (C) Both intrinsic and extrinsic effects of the metal on the complex are considered. Light is assumed unpolarized.

In Fig. 3 (panels A and B), the simulated absorption spectra of LH2 are reported with and without the effect of NT-800. In the simulation, the light is assumed unpolarized (analogous spectra with polarized light are presented in the ESI Fig. S7†). Panel A illustrates the unperturbed case, which displays the two typical absorption bands with a ∼2 : 1 ratio which reflects the fact that the B850 ring contains twice as many BChls as the B800 ring (18 and 9, respectively). Panel B shows the intrinsic effect of NT-800 on the system: while the state energies are only slightly modified, a certain redistribution of oscillator strength at the high-energy end of the spectrum is noticeable. This mainly concerns the B800 states resonant with the tip, as S10, S16 and S22 gain strength to the detriment of the bright S11 and S12 states, with a consequent increase of the B800 band width. Notice also the internal dipole redistribution affecting previously identical S11 and S12 states.

The modifications observed are not strong enough to overturn the overall shape of the spectrum, and in the end the intrinsic effect of the metal is mainly limited to breaking the original symmetry of the exciton states, rather than modifying the shape of the absorption spectra. Such perturbation of the exciton states, in the case of NT-800, is dominated by diagonal contributions (i.e., on the site energies, rather than on the couplings), but this is not necessarily true for all metal aggregates; indeed, when the NT-850 is considered, the shifts induced on the B850 site energies are comparable to those induced on the couplings, and the metal affects both diagonal and off-diagonal terms of the Hamiltonian matrix. Anyway, as the effect is now concentrated on the strongly coupled B850 ring, it turns out that the Hamiltonian is much more robust to the (weaker) modifications operated by the metal (which is also the origin of the exchange narrowing phenomenon[55]), and the marked localization effects observed for NT-800 cannot be seen with NT-850. In short, the exciton states delocalized on the B850 ring, including the bright S2 and S3 states, are virtually unaffected by the presence of the tip, while we shall later see that the explicit interaction with light mediated by the plasmonic tip (extrinsic effect) can produce a superposition of such states with a rather localized character.

## Extrinsic effect: modification of the exciton interaction with light

The extrinsic effect can be conveniently reformulated using an effective transition dipole moment, the absorption coefficients being proportional to its square modulus. Indeed, when a molecular system in a complex polarizable environment (here constituted by the protein scaffold and the metal) is subject to an external electric field, the induced transition dipole moment is $\mu^{tr} = \mu^{tr}_{sys} + \mu^{tr}_{prot} + \mu^{tr}_{met}$, where the last three terms are the transition dipole of the absorbing system, and the dipoles it induces in the polarizable protein environment and the metal nanoparticle, respectively. The latter term is in general complex, due to the dissipative dielectric response of the metal nanoparticle. If the MNP is resonant with the transition frequency, $\mu^{tr}_{met}$ can be overwhelmingly larger than the other contributions, provided the optimal shape, orientation and position are selected. In the present case, where an exciton system is at play, the magnitude of this effect is different for each of the exciton states, resulting in a radically altered absorption spectrum. Indeed, for selected exciton states, $\mu^{tr}_{met}$ can be more than one order of magnitude larger than $\mu^{tr}_{sys}$. For instance, recent single-molecule observations and computational simulations showed that the fluorescence of the LH2 system close to metal nanorods displayed enhancements of up to three orders of magnitude.[34,35,37] These results were obtained using a setup characterized by a large extrinsic metal effect, which however did not significantly perturb the exciton states, unlike the present work.

In Fig. 3C we show the spectrum simulated including the whole perturbation induced by NT-800, assuming unpolarized light: the oscillator strength is computed using the effective transition dipole moment $\mu^{tr}$, therefore comprising the extrinsic contributions from the tip. We observe that the effect of the tip is larger at the high-energy end of the spectrum, which is resonant with the plasmon frequency, and more limited for the low energy states. As expected, the opposite is true when tip NT-850, with plasmon peak at 850 nm, is analyzed instead (see ESI Fig. S11†). We also note that the loss of dipole degeneracy in the bright states is now further increased. For instance, the metal tip enhances the dipole strength of the initially degenerate bright states S11 and S12 by approximately 25 and 170 times, respectively. Indeed, the asymmetrically placed tip responds differently to the localized states, depending on the distance and orientation of their transition dipole moment with respect to the metal polarization axis. Therefore, combinations otherwise degenerate in energy and properties, like the two fingerprint pairs of bright states, induce responses that are quite different. This is discussed and exemplified in ESI Section S1.† We finally note that, as a consequence of this extrinsic effect of the metal, states previously dark or very weak can gain much strength, and therefore become accessible to light probing. Indeed, the shape of the simulated absorption spectrum in Fig. 3C is completely transformed, with the B800 band at higher energy being more than five times as intense as the B850 one. Moreover, the spectrum is dominated by the initially dark B800 state S22. As observed before, the intrinsic effect provided a strong localization for this state, with an increase in strength, which was further intensified by the extrinsic effect. Note the absolute value of the absorption spectrum, which has increased by more than three orders of magnitude with respect to the unperturbed one. Indeed, this is due to the very large extrinsic metal effect, and is responsible for the large absorption and fluorescence enhancements observed for a large variety of systems close to metal nanoparticles. Spectra simulated in the presence of NT-850, reported in the ESI,† show different enhancements.

We now investigate the practical implications of being able to populate perturbed exciton states, or combinations thereof, that were previously inaccessible. We will simulate the effect of light irradiation of the system at certain frequency ranges, thus populating the exciton states absorbing at those frequencies. Should more than one state absorb in the range considered, the effective state produced will be a superposition, to which each state $M$ with energy in the range contributes proportionally to the component of its effective dipole moment along the polarization direction of the light: $|\Psi\rangle \propto \sum_{M} (\hat{\varepsilon} \cdot \mu_M^{tr}) |M\rangle$ (see eqn (3)).

This amounts to considering that the system is investigated by ultrashort laser pulses (ideally behaving as delta-pulses), like those used in ultrafast spectroscopies.

In the absence of any perturbation, the states accessible to light probing are the two couples of bright states (S2, S3 and S11, S12), characterized by a complete delocalization over the B800 and B850 rings, respectively. The metal nanoparticle instead gives direct access to states previously inaccessible, whose spatial arrangements, in addition, may be considerably perturbed. As an example, we simulated the superposition of states obtained by irradiating the LH2 system in the presence of NT-850 or NT-800. The irradiation was simulated with light polarized along the NT axis, in energy ranges resonant with each tip, i.e., around 850 and 800 nm for NT-850 and NT-800, respectively. The results are reported in Fig. 4 (left). Both superpositions happen to be highly localized on the tip side, with effective PRs of 4.9 and 1.8 (top and bottom, respectively).

When irradiation at high energy is considered, in the presence of the resonant NT-800 (bottom panel), state S22 is dominant (40%) and the superposition is localized on BChls $\gamma_1$ and $\gamma_9$ (73% and 14%). This is a particularly interesting case as S22 is a dark state that can now be accessed thanks to the extrinsic metal effect, which increases its effective dipole moment. At the same time, however, for this B800 state the intrinsic effect is of the highest importance, as it perturbs the state character, localizing it. The resulting combination is therefore highly asymmetrical, and characterized by one large contribution from the $\gamma_1$ BChl. When instead irradiation at low energy is considered, in the presence of the resonant NT-850 (top panel), the superposition concerns mainly BChls on the B850 ring and is also rather localized on the side of LH2 closest to the NT-850 tip. This is somewhat surprising, as the tip, resonant with the low energy states, had a negligible intrinsic effect on the BChl site energies and couplings compared to the NT-800 one, as already discussed. Moreover, the B850 BChls show a much stronger inter-chromophore coupling, and are therefore much more robust to perturbations. As a consequence, the B850 exciton states were quite little modified by the intrinsic effect of the tip, with negligible variations of PRs and dipole moment redistributions. The trick here is played by the extrinsic effect, enhancing the dipole moments of states S3, S5 and S7, so that in the 1.35–1.50 eV range they contribute by 85% to the superposition. Despite the fact that neither of the three states is particularly perturbed by the NT (and thus all the states are rather delocalized, with PR $\geq 10.7$) this particular combination shows a clear localization on the $\alpha$ and $\beta$ BChls of monomers 9 (50%), 1 and 2 (16% each). Note that the cases considered, with the light polarized along the metal P axis, are those that maximally enhance the effect of the tip, since the exciton dipole moments induced on the metal are directed along $P$. Larger protein–metal distances were also tested: the extrinsic effect remains strong, while the intrinsic one decays rapidly, yielding a similar trend to what seen in ref. 37.

The results obtained so far refer to an ideal symmetric geometry, disregarding any effect arising from fluctuations. At room temperature, such results could be completely overturned, should the disorder overcome the effect of the nanoparticle. Moreover, it is known that, in the presence of static disorder, the B800 excitations are very much localized already,[56] owing to the small coupling between BChls. While this could seem to belittle the effect of the plasmon shown so far for B800 excitations, it in fact suggests that the latter could act as a selector, allowing to selectively localize the excitation onto one particular BChl. In order to verify this hypothesis, the effect of the static disorder has been modeled by adding a normally distributed noise on the site energies. The resulting excitations have been computed for 100 000 realizations with normally distributed diagonal static disorder, and the average populations are reported in Fig. 4 (right panel). Clearly, the disorder does not significantly modify the picture obtained for the static case, as the same spatial localization can be observed. The average absorption spectra, simulated using Lorentzian line-shapes with 5 cm$^{-1}$ homogeneous broadening, are shown in ESI Section S1.†

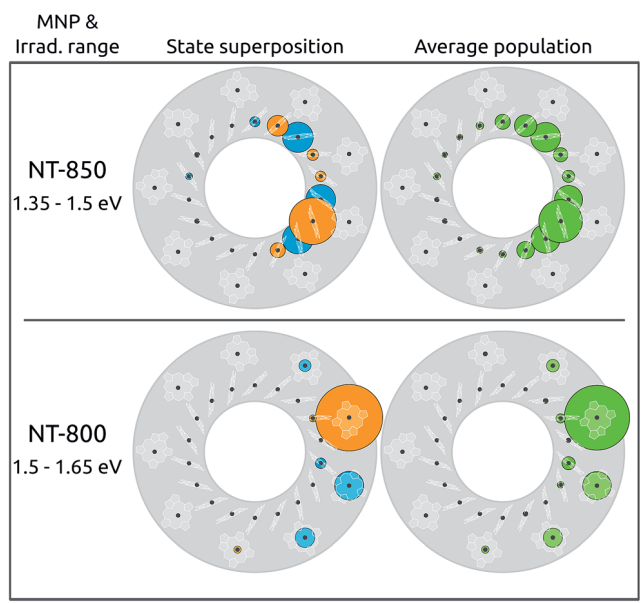

Fig. 4 Left: superposition of states accessible using light polarized along the main NT axis in the 1.35–1.5 (top) and 1.5–1.65 eV (bottom) ranges. Right: average population of each BChl under the same irradiation, over 100 000 realisations obtained by including a diagonal static disorder. The LH2 system is perturbed by NT-850 (top) and NT-800 (bottom). Significant contributions to the average population: for NT-850 BChls $\beta_9$ (31%), $\beta_2$ (15%), $\alpha_9$ (14%), $\alpha_1$ (13%); average PR: 6.1. For NT-800 BChls $\gamma_1$ (69%), $\gamma_9$ (13%); average PR: 2.0.

## Conclusion

In summary, the results shown in the present work demonstrate that it is possible to devise metal nanostructures able to drastically perturb the character of exciton states in multichromophoric systems, inducing localization effects of varying degrees. While it is well-known that the metal nanoparticles can induce huge enhancements (or alternatively quenching) in the optical spectra of light-harvesting pigment–protein complexes,[30,34,57–59] here we have shown that optimally tuned tip-shaped particles can selectively excite localized regions of typically coherent systems, eventually narrowing down to probing one single pigment. Our calculations show that this can be achieved on both B800 and B850 rings of LH2, thus opening novel interesting scenarios, as it allows to shed light on the fundamental properties of multichromophoric systems, particularly for what concerns the effect of coherences in the light capture and energy transfer processes.[27] One can envisage ultrafast experiments where the excitation is consistently prepared on a single chromophore inside the ring, and the following quantum diffusion is probed by subsequent light pulses. This would make it possible to take advantage of the distinctive properties of exciton systems (among which the long lifetimes associated with the states and the highly efficient energy transfer) but also of the ability to selectively control the pigments' excitations, in terms of energy or spatial position, thus using the metal aggregate as a selector or switch between energy and charge transfer pathways.

## Conflicts of interest

There are no conflicts to declare.

## Acknowledgements

S. Corni acknowledges funding from the ERC under the grant ERC-CoG-681285 TAME-Plasmons.

## Notes and references

1 G. C. Schatz, M. A. Young and R. P. van Duyne, in *Surface-Enhanced Raman Scattering: Physics and Applications*, ed. K. Kneipp, M. Moskovits and H. Kneipp, Springer Berlin Heidelberg, Berlin, Heidelberg, 2006, pp. 19–45.

2 S. Lal, N. K. Grady, J. Kundu, C. S. Levin, J. B. Lassiter and N. J. Halas, *Chem. Soc. Rev.*, 2008, **37**, 898–911.

3 K. A. Willets and R. P. V. Duyne, *Annu. Rev. Phys. Chem.*, 2007, **58**, 267–297.

4 V. Giannini, A. I. Fernández-Domínguez, S. C. Heck and S. A. Maier, *Chem. Rev.*, 2011, **111**, 3888–3912.

5 K. M. Mayer and J. H. Hafner, *Chem. Rev.*, 2011, **111**, 3828–3857.

6 J.-E. Park, J. Kim and J.-M. Nam, *Chem. Sci.*, 2017, **8**, 4696–4704.

7 T. Tatsuma, H. Nishi and T. Ishida, *Chem. Sci.*, 2017, **8**, 3325–3337.

8 K. Kneipp, H. Kneipp, I. Itzkan, R. R. Dasari and M. S. Feld, *Chem. Rev.*, 1999, **99**, 2957–2976.

9 M. D. Sonntag, J. M. Klingsporn, A. B. Zrimsek, B. Sharma, L. K. Ruvuna and R. P. Van Duyne, *Chem. Soc. Rev.*, 2014, **43**, 1230–1247.

10 P. Anger, P. Bharadwaj and L. Novotny, *Phys. Rev. Lett.*, 2006, **96**, 113002.

11 S. Kühn, U. Håkanson, L. Rogobete and V. Sandoghdar, *Phys. Rev. Lett.*, 2006, **97**, 1–4.

12 R. Zhang, Y. Zhang, Z. C. Dong, S. Jiang, C. Zhang, L. G. Chen, L. Zhang, Y. Liao, J. Aizpurua, Y. Luo, J. L. Yang and J. G. Hou, *Nature*, 2013, **498**, 82–86.

13 Y. Sugawara, T. A. Kelf, J. J. Baumberg, M. E. Abdelsalam and P. N. Bartlett, *Phys. Rev. Lett.*, 2006, **97**, 266808.

14 T. Schwartz, J. A. Hutchison, C. Genet and T. W. Ebbesen, *Phys. Rev. Lett.*, 2011, **106**, 196405.

15 T. W. Ebbesen, *Acc. Chem. Res.*, 2016, **49**, 2403–2412.

16 P. Vasa and C. Lienau, *ACS Photonics*, 2018, **5**, 2–23.

17 R. Chikkaraddy, B. De Nijs, F. Benz, S. J. Barrow, O. A. Scherman, E. Rosta, A. Demetriadou, P. Fox, O. Hess and J. J. Baumberg, *Nature*, 2016, **535**, 127–130.

18 S. Duan, G. Tian, Y. Ji, J. Shao, Z. Dong and Y. Luo, *J. Am. Chem. Soc.*, 2015, **137**, 9515–9518.

19 P. Liu, D. V. Chulhai and L. Jensen, *ACS Nano*, 2017, **11**, 5094–5102.

20 F. Benz, M. K. Schmidt, A. Dreismann, R. Chikkaraddy, Y. Zhang, A. Demetriadou, C. Carnegie, H. Ohadi, B. de Nijs, R. Esteban, J. Aizpurua and J. J. Baumberg, *Science*, 2016, **354**, 726.

21 P. Vasa, C. Ropers, R. Pomraenke and C. Lienau, *Laser Photonics Rev.*, 2009, **3**, 483–507.

22 D. Brinks, M. Castro-lopez, R. Hildner and N. F. V. Hulst, *Proc. Natl. Acad. Sci. U. S. A.*, 2013, **110**, 18386–18390.

23 N. L. Gruenke, M. F. Cardinal, M. O. McAnally, R. R. Frontiera, G. C. Schatz and R. P. Van Duyne, *Chem. Soc. Rev.*, 2016, **45**, 2263–2290.

24 L. Piatkowski, N. Accanto and N. F. Van Hulst, *ACS Photonics*, 2016, **3**, 1401–1414.

25 R. J. Cogdell, A. M. Hawthornthwaite, M. B. Evans, L. A. Ferguson, K. Kerfeld, J. Thornber, F. van Mourik and R. van Grondelle, *Biochim. Biophys. Acta, Bioenerg.*, 1990, **1019**, 239–244.

26 S. K. Saikin, A. Eisfeld, S. Valleau and A. Aspuru-Guzik, *Nanophotonics*, 2013, **2**, 21–38.

27 G. D. Scholes, G. R. Fleming, L. X. Chen, A. Aspuru-Guzik, A. Buchleitner, D. F. Coker, G. S. Engel, R. van Grondelle, A. Ishizaki, D. M. Jonas, J. S. Lundeen, J. K. McCusker, S. Mukamel, J. P. Ogilvie, A. Olaya-Castro, M. A. Ratner, F. C. Spano, K. B. Whaley and X. Zhu, *Nature*, 2017, **543**, 647–656.

28 D. M. Wilkins and N. S. Dattani, *J. Chem. Theory Comput.*, 2015, **11**, 3411–3419.

29 S. M. Blau, D. I. G. Bennett, C. Kreisbeck, G. D. Scholes and A. Aspuru-Guzik, ArXiv e-prints, 2017.

30 S. Mackowski, S. Wörmke, A. J. Maier, T. H. P. Brotosudarmo, H. Harutyunyan, A. Hartschuh,

A. O. Govorov, H. Scheer and C. Bräuchle, *Nano Lett.*, 2008, **8**, 558–564.

31 M. Cohen, O. Heifler, Y. Lilach, Z. Zalevsky, V. Mujica, I. Carmeli and S. Richter, *Nat. Commun.*, 2015, **6**, 1–6.

32 A. Tsargorodska, M. L. Cartron, C. Vasilev, G. Kodali, O. A. Mass, J. J. Baumberg, P. L. Dutton, C. N. Hunter, P. Törmä and G. J. Leggett, *Nano Lett.*, 2016, **16**, 6850–6856.

33 S. R. Beyer, S. Ullrich, S. Kudera, A. T. Gardiner, R. J. Cogdell and J. Köhler, *Nano Lett.*, 2011, **11**, 4897–4901.

34 E. Wientjes, J. Renger, A. G. Curto, R. Cogdell and N. F. van Hulst, *Nat. Commun.*, 2014, **5**, 4236.

35 E. Wientjes, J. Renger, A. G. Curto, R. Cogdell and N. F. van Hulst, *Phys. Chem. Chem. Phys.*, 2014, **16**, 24739–24746.

36 O. Andreussi, A. Biancardi, S. Corni and B. Mennucci, *Nano Lett.*, 2013, **13**, 4475–4484.

37 S. Caprasecca, C. A. Guido and B. Mennucci, *J. Phys. Chem. Lett.*, 2016, **7**, 2189–2196.

38 M. Z. Papiz, S. M. Prince, T. Howard, R. J. Cogdell and N. W. Isaacs, *J. Mol. Biol.*, 2003, **326**, 1523–1538.

39 D. Case, J. Berryman, R. Betz, D. Cerutti, T. Cheatham III, T. Darden, T. Giese, H. Gohlke, A. Goetz, N. Homeyer, S. Izadi, P. Janowski, J. Kaus, A. Kovalenko, T. Lee, S. LeGrand, P. Li, T. Luchko, R. Luo, B. Madej, K. Merz, G. Monard, P. Needham, H. Nguyen, H. Nguyen, I. Omelyan, A. Onufriev, D. Roe, A. Roitberg, R. Salomon-Ferrer, C. Simmerling, W. Smith, J. Swails, R. Walker, J. Wang, R. Wolf, X. Wu, W. Zhang, G. Seabra, K. Wong, F. Paesani, J. Vanicek, R. Wolf, J. Liu, X. Wu, D. York and P. Kollman, *Amber14*, University of California, San Francisco, 2015.

40 L. W. Chung, H. Hirao, X. Li and K. Morokuma, *Wiley Interdiscip. Rev.: Comput. Mol. Sci.*, 2012, **2**, 327–350.

41 L. Cupellini, S. Jurinovich, M. Campetella, S. Caprasecca, C. A. Guido, S. M. Kelly, A. T. Gardiner, R. Cogdell and B. Mennucci, *J. Phys. Chem. B*, 2016, **120**, 11348–11359.

42 C. Curutchet, A. Muñoz Losa, S. Monti, J. Kongsted, G. D. Scholes and B. Mennucci, *J. Chem. Theory Comput.*, 2009, **5**, 1838–1848.

43 S. Caprasecca, C. Curutchet and B. Mennucci, *PolChat: A polarisation-consistent charge-fitting tool*, University of Pisa, 2015, http://molecolab.dcci.unipi.it/tools, Molecolab Tools.

44 J. Tomasi, B. Mennucci and R. Cammi, *Chem. Rev.*, 2005, **105**, 2999–3094.

45 S. Corni and J. Tomasi, *J. Chem. Phys.*, 2002, **117**, 7266.

46 S. Corni and J. Tomasi, *J. Chem. Phys.*, 2003, **118**, 6481.

47 O. Andreussi, S. Corni, B. Mennucci and J. Tomasi, *J. Chem. Phys.*, 2004, **121**, 10190.

48 D. S. Indrasekara, A. Swarnapali, R. Thomas and L. Fabris, *Phys. Chem. Chem. Phys.*, 2015, **17**, 21133–21142.

49 *Handbook of Optical Constants of Solids*, ed. E. D. Palik, Academic Press, New York, 1985.

50 A. Angioni, S. Corni and B. Mennucci, *Phys. Chem. Chem. Phys.*, 2013, **15**, 3294–3303.

51 D. Thouless, *Phys. Rep.*, 1974, **13**, 93–142.

52 F. Segatta, L. Cupellini, S. Jurinovich, S. Mukamel, M. Dapor, S. Taioli, M. Garavelli and B. Mennucci, *J. Am. Chem. Soc.*, 2017, **8**, 2357–2367.

53 L.-Y. Hsu, W. Ding and G. C. Schatz, *J. Phys. Chem. Lett.*, 2017, **139**, 7558–7567.

54 M. J. Frisch, G. W. Trucks, H. B. Schlegel, G. E. Scuseria, M. A. Robb, J. R. Cheeseman, G. Scalmani, V. Barone, B. Mennucci, G. A. Petersson, H. Nakatsuji, M. Caricato, X. Li, H. P. Hratchian, A. F. Izmaylov, J. Bloino, G. Zheng, J. L. Sonnenberg, M. Hada, M. Ehara, K. Toyota, R. Fukuda, J. Hasegawa, M. Ishida, T. Nakajima, Y. Honda, O. Kitao, H. Nakai, T. Vreven, J. A. Montgomery Jr, J. E. Peralta, F. Ogliaro, M. Bearpark, J. J. Heyd, E. Brothers, K. N. Kudin, V. N. Staroverov, R. Kobayashi, J. Normand, K. Raghavachari, A. Rendell, J. C. Burant, S. S. Iyengar, J. Tomasi, M. Cossi, N. Rega, J. M. Millam, M. Klene, J. E. Knox, J. B. Cross, V. Bakken, C. Adamo, J. Jaramillo, R. Gomperts, R. E. Stratmann, O. Yazyev, A. J. Austin, R. Cammi, C. Pomelli, J. W. Ochterski, R. L. Martin, K. Morokuma, V. G. Zakrzewski, G. A. Voth, P. Salvador, J. J. Dannenberg, S. Dapprich, A. D. Daniels, Â. Farkas, J. B. Foreman, J. V. Ortiz, J. Cioslowski and D. J. Fox, *Gaussian09 Revision D.01*, Gaussian Inc., Wallingford CT, 2009.

55 E. Knapp, *Chem. Phys.*, 1984, **85**, 73–82.

56 Y. C. Cheng and R. J. Silbey, *Phys. Rev. Lett.*, 2006, **96**, 028103.

57 I. Kim, S. L. Bender, J. Hranisavljevic, L. M. Utschig, L. Huang, G. P. Wiederrecht and D. M. Tiede, *Nano Lett.*, 2011, **11**, 3091–3098.

58 S. R. Beyer, S. Ullrich, S. Kudera, A. T. Gardiner, R. J. Cogdell and J. Köhler, *Nano Lett.*, 2011, **11**, 4897–4901.

59 M. Szalkowski, J. D. Janna Olmos, D. Buczynska, S. Mackowski, D. Kowalska and J. Kargul, *Nanoscale*, 2017, **9**, 10475–10486.